# Free-hand gas identification based on transfer function ratios without gas flow control

Gaku Imamura[1,2]*, Kota Shiba[1,2], Genki Yoshikawa[1,2,3], and Takashi Washio[4]

[1]*World Premier International Research Center Initiative (WPI), International Center for Materials Nanoarchitectonics (MANA), National Institute for Materials Science (NIMS), Tsukuba, Ibaraki, 305-0044, Japan*
[2]*Center for Functional Sensor & Actuator (CFSN), National Institute for Materials Science (NIMS), Tsukuba, Ibaraki, 305-0044, Japan*
[3]*Materials Science and Engineering, Graduate School of Pure and Applied Science, University of Tsukuba, Tennodai 1-1-1 Tsukuba, Ibaraki 305-8571, Japan*
[4]*The Institute of Scientific and Industrial Research, Osaka University, Mihogaoka 8-1, Ibaraki, Osaka 567-0047. Japan*
*E-mail: IMAMURA.Gaku@nims.go.jp

## Abstract

Gas identification is one of the most important functions of gas sensor systems. To identify gas species from sensing signals, however, gas input patterns (*e.g.* the gas flow sequence) must be controlled or monitored precisely with additional instruments such as pumps or mass flow controllers; otherwise, effective signal features for analysis are difficult to be extracted. Toward a compact and easy-to-use gas sensor system that can identify gas species, it is necessary to overcome such restrictions on gas input patterns. Here we develop a novel gas identification protocol that is applicable to arbitrary gas input patterns without controlling or monitoring any gas flow. By combining the protocol with newly developed MEMS-based sensors—Membrane-type Surface stress Sensors (MSS), we have realized the gas identification with the free-hand measurement, in which one can simply hold a small sensor chip near samples. From sensing signals obtained through the free-hand measurement, we have developed machine learning models that can identify not only solvent vapors but also odors of spices and herbs with high accuracies. Since no bulky gas flow control units are required, this protocol will expand the applicability of gas sensors to portable electronics and wearable devices, leading to practical artificial olfaction.

Introduction

Recent advances in information and communication technology (ICT) have stimulated huge demand for sensors, which play a key role in highly integrated systems—for example, cyber-physical systems (CPS) and Internet of things (IoT) [1]. Among sensors, gas sensors have been used in various applications such as detection of toxic gases, monitor of indoor air quality, and automotive emissions control [2]. In addition to these applications which basically focus on single component gases, development of a system that can detect and identify odors—complex mixture of gases—has been a long-standing issue since Persaud and Dodd published the first report on artificial olfaction based on gas sensors in 1982 [3]. The basic concept of artificial olfaction is as follows: first, an odor is detected with an array of gas sensors, each of which shows a different sensing property. Then, the signal features are extracted and/or selected from the sensing signals. Finally, the odor is identified on the basis of the features through a classification algorithm. Therefore, it is necessary to develop both gas sensors and data analysis methods to realize artificial olfaction. Many studies on artificial olfaction have been published so far [4,5], and some products are already commercially available. However, products that meet the requirements for consumer use have not appeared yet; odor identification devices which are low-priced, portable, and easy-to-use have not been achieved. To realize such a practical artificial olfaction, it is essential to develop a simple and compact measurement system that people without specific expertise can use.

For realizing practical artificial olfaction, gas flow control is one of the biggest problems. To obtain comparable sensing signals, it is required to employ the same gas flow sequence for every measurement [6]. Thus, in many odor measurement systems, a sample gas is injected to a gas sensor by gas flow control units such as pumps and mass flow controllers (MFCs). Typically, a sample gas and a carrier gas are alternatively injected to a gas sensor, resulting in a periodic peaks of sensing responses. From such sensing signals, features such as slope, area, and decay time are extracted for analysis. Since the shape of sensing signals strongly depends on the gas flow sequence in this measurement protocol, measurement data obtained with a different gas flow sequence cannot be compared with each other. To overcome this issue, gas identification methods based on system identification have been developed. A gas sensing system can be regarded as an input-output system; gas injection patterns (*e.g.* the gas flow control or the gas concentration) and sensing signals are inputs and outputs, respectively. In this input-output system, the sensing response can be analyzed through various system identification methods. In 1994, Nakamura *et al.* demonstrated that the sensing response of quartz crystal microbalance (QCM) can be analyzed by autoregressive (AR) models [7], leading to the development of analysis with advanced time-series models such as autoregressive models with exogenous input (ARX) and autoregressive moving average (ARMA) models [8,9]. Analysis methods based on an impulse response function and a transfer function were also developed to describe the dynamic behavior of sensor responses to varying gas concentration [8,10,11]. Furthermore, pioneering work was done by Marco and Pardo and their coworkers to adapt non-linear models including artificial neural networks (ANN) for describing the complex response of sensing systems [12-14]. By utilizing these system identification methods, it became possible to estimate the process model of a sensing system independently of gas input patterns. However, such analysis methods based on system identification require the information of gas input patterns; otherwise the relationship between inputs and outputs cannot be estimated. Thus, pumps or gas flow monitors are still needed for a measurement system to obtain temporal changes of inputs.

Toward simple and flexible artificial olfaction, a measurement protocol that requires neither control nor monitor of the gas input pattern is required. Several groups have reported such measurement protocols; that is, the gas identification methods for an open environment without gas flow control. Trincavelli and his colleagues developed a gas identification system with continuous sampling, in which the system continuously intakes a plume of a sample gas [15-18]. The authors focused on the transient information in the sensing signals and demonstrated the feasibility of the method by using a mobile robot equipped with an electronic nose. Vergara *et al.* investigated the classification performance of gas sensor arrays by using a wind tunnel testbed facility where gas sensor arrays are exposed to sample gases, while the authors used the response of each gas sensor at a steady state as features for classification [19]. Related to these works, Han *et al.* developed an improved classification algorithm based on an unsupervised learning—the *KmP* algorithm [20]. Although their studies using metal oxide semiconductor (MOS) gas sensors have exploited gas identification protocol without gas flow control, analysis methods used in these studies are adaptable to gas input patterns which do not vary significantly for every measurement. In other words, these methods are valid only when gas input patterns are statistically stable. Toward a practical application of artificial olfaction, further breakthrough is still needed to enhance the usability of the measurement system.

In this study, we have developed a novel gas identification protocol which requires neither control nor monitor of gas input patterns; gas species can be identified only from the transient responses of arrayed gas sensors without using pumps or MFCs. Combined with a miniaturized sensor including an MEMS sensor, this gas identification protocol realizes a compact measurement system in which gas species are identified through the *free-hand measurement*—sample gases are measured with a small sensor chip by manually moving the sensor chip near the sample (Fig. 1a). The key to this identification protocol is a novel data analysis method focusing on the transient responses of arrayed sensors. By comparing the sensing responses of two different channels, the transfer function ratio (TFR) can be estimated without using gas input patterns as explained later in detail. Since the TFR is intrinsic to the combination of sensors and gas species, the gas species can be identified by TFR. To demonstrate the gas identification through the free-hand measurement, we employed Membrane-type Surface stress Sensors (MSS) as gas sensors in this study because of their high sensitivity, compactness, and wide variation in sensing properties (Fig. 1b) [21,22]. From the measurement data obtained through the free-hand measurement (Fig. 1c), we developed machine learning models for classifying gas species, resulting in identification of four solvent vapors at an accuracy of 0.997±0.004. Furthermore, we applied this gas identification protocol to odors of spices and herbs, resulting in a classification accuracy of 0.89±0.02. This result shows that not only single component gases but also multicomponent gas mixtures can be identified simply by moving a small sensor chip near samples. The robustness of the TFR-based analysis to gas input patterns was confirmed by developing a classification model from measurement data obtained with two different types of gas injection sequences. Based on this TFR-based gas identification protocol combined with an MEMS sensor chip, artificial olfaction can be implemented in various portable devices and wearable devices.

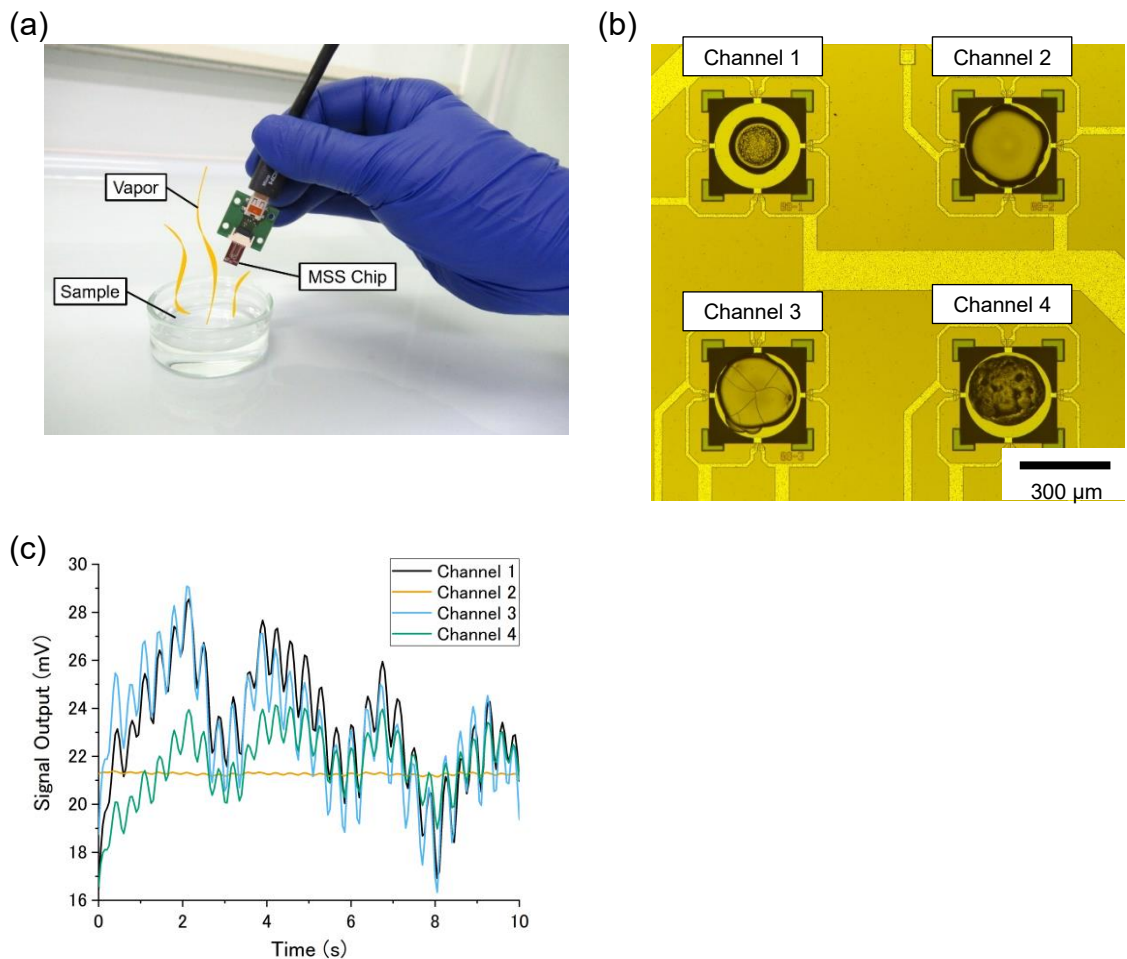

**Figure 1**   (a) The picture of the free-hand measurement with the illustration of sample vapor. (b) The optical microscope image of MSS Chip I. Channels 1 to 4 are coated with poly(vinylidene fluoride), polysulfone, poly(4-methylstyrene) and polycaprolactone, respectively. (c) Sensing signals of MSS Chip I measuring ethyl acetate through the free-hand measurement.

Proposed Gas Identification Protocol

A transfer function is one of the mathematical representations to describe a process model of a system. Assuming that a gas sensing system exhibits linear response, in which an output sensing signal $y(t)$ is linear in the gas injection pattern $x(t)$, $y(t)$ can be described as a convolution of $x(t)$ and the time-domain transfer function (or the impulse response function) $h_g(t)$:

$$y(t) = \int_0^t h_g(\tau) x(t-\tau) d\tau \tag{1}$$

Here, $h_g(t)$ is determined by the interaction between the sensor and a gas $g$. As $h_g(t)$ does not depend on $x(t)$, $h_g(t)$ is considered to be an intrinsic function of the gas, leading to gas identification based on $h_g(t)$. By applying the Fourier transform, the frequency-domain expression for Equation (1) can be obtained as the following form:

$$Y(f) = H_g(f) X(f) \tag{2}$$

where $X(f)$, $Y(f)$, and $H_g(f)$ are the frequency-domain expressions for the gas injection pattern, the output sensing signal, and the transfer function, respectively[8].

It is, of course, possible to identify gas species by the transfer function $H_g(f)$ calculated from gas input pattern $X(f)$ (e.g. gas flow rate or gas concentration) and sensing signals $Y(f)$. In such a straightforward approach, however, $X(f)$ still needs to be measured by controlling or monitoring gas injections to calculate $H_g(f)$. This issue can be solved by using an array of gas sensors with different sensing characteristics. Considering that a gas $g$ is provided to an array of gas sensors according to $X(f)$, the output sensing signal of the $i$th channel of the sensor array $Y_i(f)$ is described as the following form:

$$Y_i(f) = H_{g,i}(f) X(f) \tag{3}$$

where $H_{g,i}(f)$ is the transfer function of the $i$th channel to the gas $g$. If gas sensors in the array can be considered to be spatially equivalent for the gas input, $X(f)$ is the same for all the channels. Thus, for any combination of two channels $m$ and $n$, the following equation holds:

$$X(f) = \frac{Y_m(f)}{H_{g,m}(f)} = \frac{Y_n(f)}{H_{g,n}(f)} \tag{4}$$

Let $K_{m,n}(f)$ be defined as $K_{m,n}(f) = Y_m(f)/Y_n(f)$, which is the signal ratio of the $m$th and $n$th channels in the frequency domain. Then, $K_{m,n}(f)$ can be described as the following form from Equation (4):

$$K_{m,n}(f) = \frac{Y_m(f)}{Y_n(f)} = \frac{H_{g,m}(f)}{H_{g,n}(f)} \tag{5}$$

As $H_{g,m}(f)/H_{g,n}(f)$ is the TFR of $m$th and $n$th channels, $K_{m,n}(f)$ is the intrinsic value to the gas $g$. Here, it is noteworthy that $K_{m,n}(f)$ is independent of $X(f)$; that is, $K_{m,n}(f)$ can be estimated from any gas input pattern. Therefore, by calculating $K_{m,n}(f)$ from an arbitrary combination of two channels from a gas sensor array, it is possible to identify a gas species without controlling or monitoring the gas input pattern.

Experimental Setup

Based on the gas identification protocol focusing on the TFR, it is possible to identify gas species with only a gas sensor array; gas flow lines including pumps or MFCs are no longer needed as long as a time-varying gas input that is consistent for all sensor channels is provided. Thus, in this study, we demonstrate gas identification through the free-hand measurement—sample gases are measured by manually moving a miniaturized gas sensor array near the samples. For this purpose, we utilized MSS as a sensing platform. An MSS is a kind of nanomechanical sensors, which detect changes in mechanical properties such as mass, stress, and deformation [23]. An MSS detects surface stress associated with gas sorption or desorption at a receptor layer. Its unique structure—a silicon membrane coated with a receptor material and suspended by four beams in which piezoresistors are embedded—effectively transduces the surface stress into electrical signals. MSS are suitable for the free-hand measurement owing to the following reasons. Firstly, the sensing elements (membranes) can be densely arrayed: more than 100 elements/cm$^2$ [24]. As the gas identification protocol assumes that all the channels in an array are spatially equivalent, gas sensors must be miniaturized so that all the channels are arrayed in a small area. Secondly, MSS realize various sensing characteristics, which are preferable to obtain unique $K_{m,n}(f)$. As MSS detect surface stress caused by gas sorption/desorption of the receptor material, almost all solid materials can be utilized as receptor materials of MSS, leading to a wide variety of sensing characteristics [25-27]. Thus, we developed an MSS-based measurement system, which allows the free-hand measurement without gas flow control units such as pumps and MFCs. From the data obtained through the free-hand measurements, machine learning models for gas identification were developed.

Results and Discussion

Free-hand measurement system

In this study, we used MSS chips with four channels, which were coated with different receptor materials. As receptor materials, we utilized polymers, which have been widely used in the field of nanomechanical sensors because of their preferable mechanical properties and the wide variation [28,29]. In this study, we coated the channels of an MSS chip with poly(vinylidene fluoride), polysulfone, poly(4-methylstyrene), and polycaprolactone by inkjet spotting (Fig. 1b). (hereafter, the polymer-coated MSS chip is described as "MSS Chip I".) To demonstrate gas identification with the free-hand measurement, we measured vapors of four solvents: ethanol, water, heptane, and ethyl acetate. The gas measurements were conducted by manually moving the MSS chip in the vapor of samples as shown in Fig. 1a. We measured each sample eight times for 90 seconds each time. Figure

1c shows one example of the sensing signals obtained with MSS Chip I. Each channel shows a different sensing response (*e.g.* amplitude and phase shift) according to the fluctuation of gas concentration at the chip mainly associated with a motion of a hand, which moved at roughly 3 Hz.

Using MSS Chip I, we measured the headspace vapors of the solvents through the free-hand measurement. From the measurement data, we calculated $K_{m,n}(f)$ and created a dataset for analysis. To visualize the dataset and verify the potential of $K_{m,n}(f)$ as a feature for gas identification, we first performed a typical dimensionality reduction algorithm on the dataset: principal component analysis (PCA). PCA projects data points to a low dimensional space consisting of principal components, which are determined by the variance of the dataset. Figures 2a-c show the results of PCA. Although the gas species are not completely discriminated in the plots, the data points belonging to the same gas species form a cluster on the feature spaces. As the clusters are roughly separated from each other, $K_{m,n}(f)$ reflects the different interaction between the receptor materials and the gas species.

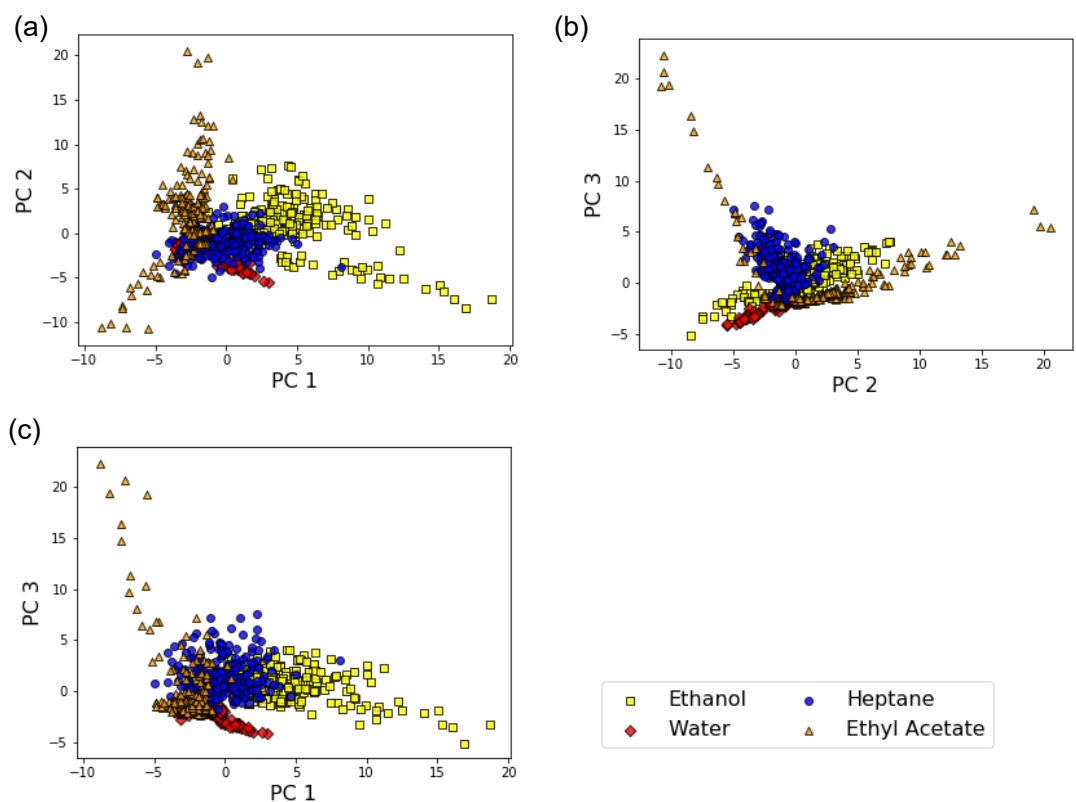

**Figure 2** The scatter plots of PCA (a-c) on the dataset of the solvent vapors obtained with MSS Chip I.

As the results of PCA show the potential of TFRs as descriptors of gas species, we then developed machine learning models for identifying the solvent vapors. The results of the models are summarized in Table 1. We achieved classification accuracies of approximately 0.85 with these classifiers, particularly the classification model based on random forests exhibits the highest accuracy among the six models (0.91±0.15). These results indicate that gas species could be identified with high accuracy by the free-hand measurement—simply moving the MSS chip in the vapor of the samples without any gas flow control. It should be noted that the measurement time required to identify gas species is only 3.0 seconds, demonstrating a rapid (practically real time) gas identification.

Table 1. Results of classification models for solvent vapors using MSS Chip I.

| Classifier | Optimized Parameters | Accuracy |
|---|---|---|
| **Support Vector Machine (linear kernel)** | Number of PCs: 20<br>$C$: 1.0 | 0.84±0.14 |
| **Support Vector Machine (RBF Kernel)** | Number of PCs: 20<br>$C$: 100.0<br>$\gamma$: 0.01 | 0.85±0.17 |
| **Logistic Regression** | Number of PCs: 20<br>$C$: 1.0 | 0.83±0.15 |
| **Decision Tree** | Maximum depth: 5 | 0.84±0.13 |
| **Random Forest** | Number of estimators: 512 | 0.91±0.15 |
| **Multilayer Perceptron** | $\alpha$: 1.0<br>Hidden layer size: (128,128,64) | 0.85±0.14 |

Optimization of receptor materials

To improve the accuracy, we optimized the receptor layers of the MSS. As an effective platform of receptor materials, we employed functional inorganic nano-/micro-particles. The nano-/micro-particles have several advantages over polymer. For example, their high surface area to volume ratio leads to high sensitivity and short response time. In the case of nanomechanical sensing, high sensitivity can be expected for inorganic nano-/micro-particles owing to their high Young's moduli [30,31]. In addition, the surface of the inorganic nano-/micro-particles can be functionalized with various kinds of moieties, providing wide variation of chemical selectivity. To enhance the chemical selectivity without deteriorating the sensitivity compared to MSS Chip I, we utilized three kinds of functional silica/titania hybrid nanoparticles (STNPs) and one kind of hybrid particles as receptor materials: C18-STNPs, Ph-STNPs, NH2-STNPs and silica-hexadecyltrimethylammonium hybrid particles (silica-C16TA hybrid). The STNPs were synthesized via sol-gel reaction of two alkoxides such as titanium tetraisopropoxide and silane coupling reagent [32,33]. The details of the synthesis method for these functional STNPs are described in References 32 and 33.The silica-C16TA hybrid was synthesized by the Stöber method combined with the supramolecular templating approach reported previously [34]. The STNPs were coated on the channels of an MSS chip through the same procedure as MSS Chip I. (hereafter "MSS Chip II".) Figure 3 shows the optical microscope image of MSS Chip II. The intensities of MSS Chip I/II to the four solvent vapors are summarized in Fig. 4. As expected, MSS

Chip II exhibited larger variation in the intensities to each vapor than MSS Chip I, indicating that MSS Chip II is more capable of discriminating the vapors than MSS Chip I.

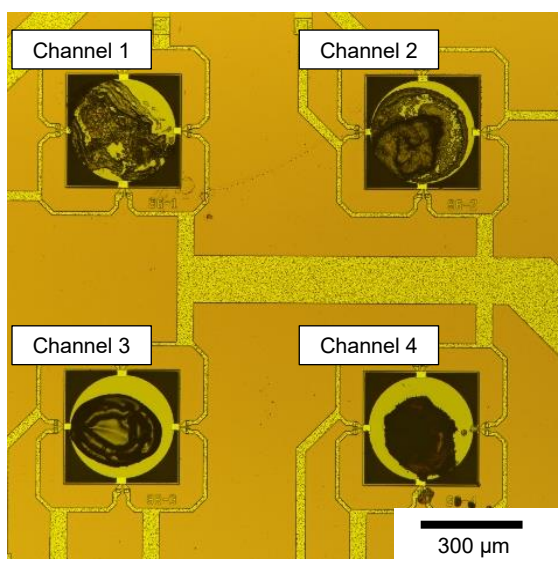

**Figure 3** The optical microscope image of MSS Chip II. Channels 1 to 4 are coated with C18-STNPs, Ph-STNPs, NH2-STNPs and silica-C16TA hybrid particles, respectively.

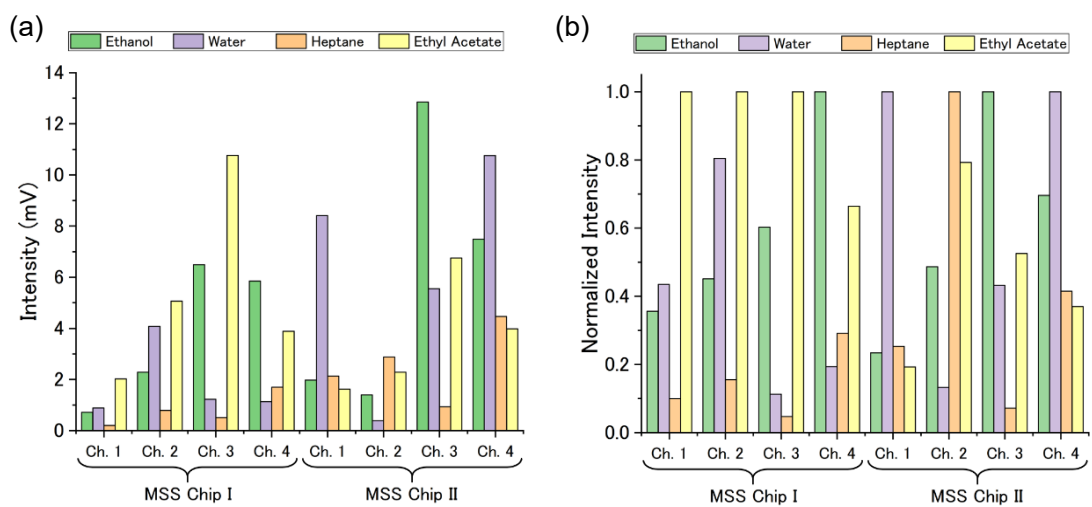

**Figure 4** (a) The intensities of MSS Chip I and MSS Chip II to the solvent vapors. (b) The intensities normalized by the highest response for each channel.

Using MSS Chip II, we conducted the free-hand measurement for the four solvent vapors. Figure 5a-c shows the results of PCA from the data obtained with MSS Chip II. The clusters appearing in Fig. 5 seem to be more separated from each other than those in Fig. 2. To quantitatively evaluate the cluster quality of the results in Figs. 2 and 5, we calculated the Davies–Bouldin (DB) index—a common index that estimates the cluster separation—for the datasets obtained with MSS Chip I and MSS Chip II. The DB index for a dataset consisting of $n$ clusters is defined as the following formula:

$$DB = \frac{1}{n}\sum_{i}^{n} \max_{i \neq j}\left(\frac{\sigma_i + \sigma_j}{d(c_i, c_j)}\right) \qquad (6)$$

where $c_k$ and $\sigma_k$ are the centroid of cluster $k$ and the mean distance of all elements in cluster $k$ from $c_k$, respectively. Distance between $c_k$ and $c_l$ is denoted as $d(c_k, c_l)$. As can be inferred from Equation (6), a DB index becomes low when clusters are well separated from each other; that is, the radius of each cluster is small, and the distance between two clusters is large. The *DB*s for the datasets obtained from MSS Chip I and MSS Chip II are 3.34 and 1.53, respectively. Therefore, the optimization of the receptor materials based on the chemical selectivity resulted in the improvement in cluster quality of the measurement dataset, leading to a good gas discrimination.

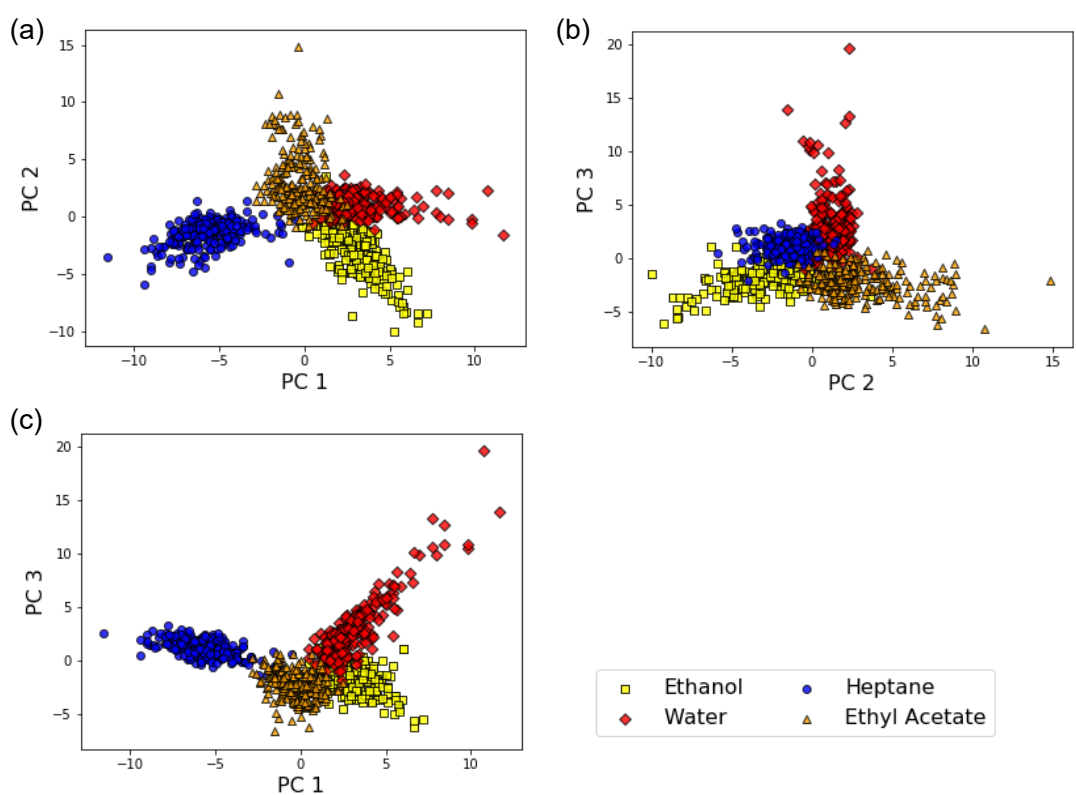

**Figure 5** The scatter plots of PCA on the dataset of the solvent vapors obtained with MSS Chip II.

The accuracy of the machine learning models was also improved by the optimization. Table 2 summarizes the details of the developed machine learning models. Compared to the previous results in Table 1, all the models developed from the data obtained with MSS Chip II show high accuracies over 0.95. Among the models, the random forest-based model achieved considerably high accuracy (0.997±0.004). It is noteworthy that the standard deviation of the accuracy is decreased for MSS Chip II, indicating that measurement data with MSS Chip II are more reliable than those with MSS Chip I.

Table 2. Results of classification models for solvent vapors using MSS Chip II.

| Classifier | Optimized Parameters | Accuracy |
|---|---|---|
| **Support Vector Machine (linear kernel)** | Number of PCs: 80<br>C: 100.0 | 0.976±0.011 |
| **Support Vector Machine (RBF Kernel)** | Number of PCs: 20<br>C: 10.0<br>γ: 0.01 | 0.977±0.011 |
| **Logistic Regression** | Number of PCs: 20<br>C: 10.0 | 0.968±0.015 |
| **Decision Tree** | Maximum depth: 5 | 0.978±0.005 |
| **Random Forest** | Number of estimators: 512 | 0.997±0.004 |
| **Multilayer Perceptron** | α: 0.1<br>Hidden layer size: (128,128,64) | 0.978±0.016 |

Odor identification of spices through the free-hand measurement

Not only vapors of solvents (single component gases) but also odors (multicomponent gas mixtures) are within the scope of this new gas identification protocol. To demonstrate odor identification through the free-hand measurement, we chose three spices as samples: rosemary, red chili pepper, and garlic. The odors of the spices were measured with MSS Chip II through the same experimental process as in the case of solvent vapors. The dataset of $K_{m,n}(f)$ was created from the measurement data and analyzed by PCA. Figure 6a-c shows the scatter plots of PCA. The formation of clusters on the plots indicates that odors can be discriminated by $K_{m,n}(f)$. The results of the developed machine learning models are summarized in Table 3. As with the case of solvent vapors, the random forest-based model exceeds other models in accuracy (0.89±0.02). These results demonstrate the odor identification with the compact measurement system consisting of only an MSS chip and electrical readout devices without any gas flow control.

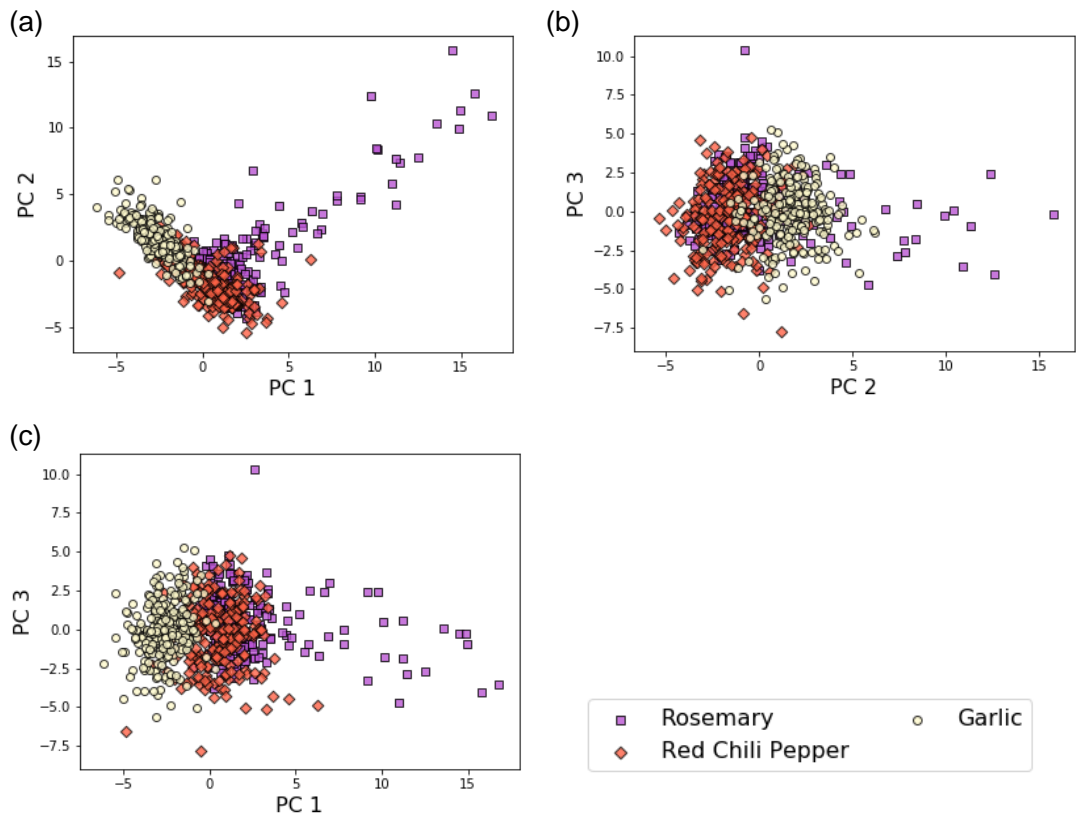

**Figure 6** The scatter plots of PCA on the dataset of the odors of the spices and herbs obtained with MSS Chip II.

Table 3. Results of classification models for spices using MSS Chip II.

| Classifier | Optimized Parameters | Accuracy |
| --- | --- | --- |
| **Support Vector Machine (linear kernel)** | Number of PCs: 40<br>$C$: 100.0 | 0.83±0.03 |
| **Support Vector Machine (RBF Kernel)** | Number of PCs: 40<br>$C$: 100.0<br>$\gamma$: 0.001 | 0.84±0.04 |
| **Logistic Regression** | Number of PCs: 40<br>$C$: 1.0 | 0.84±0.04 |
| **Decision Tree** | Maximum depth: 5 | 0.78±0.04 |
| **Random Forest** | Number of estimators: 64 | 0.89±0.02 |
| **Multilayer Perceptron** | $\alpha$: 10<br>Hidden layer size: (128,64,32) | 0.84±0.04 |

Gas identification with different gas input patterns

Finally, we conducted gas sensing measurements with different types of gas flow sequences and developed classification models from the measurement data in order to explicitly show that TFR is adaptable to any gas input pattern. Headspace gases of the four solvent vapors were injected to MSS Chip II. To generate definitely different gas input patterns, a gas flow line equipped with MFCs was utilized. Two different gas flow sequences were employed for the gas measurements: the m-sequence pseudorandom sequence (Fig. 7a) and the rectangular sequence (Fig. 7b). The sensing data obtained from the m-sequence pseudorandom sequence and the rectangular sequence were used for training and testing classification models, respectively. Note that the monitored gas flow rates were not use for analysis. Logistic regression was employed as a classifier. Figure 8a shows the results. The classification accuracy for training and test data are plotted against the regularization parameter $C$. For $C > 10^{-1}$, classification models which show high accuracy for both training and test data are developed. Therefore, it is demonstrated that gas species can be identified independently of the gas input pattern.

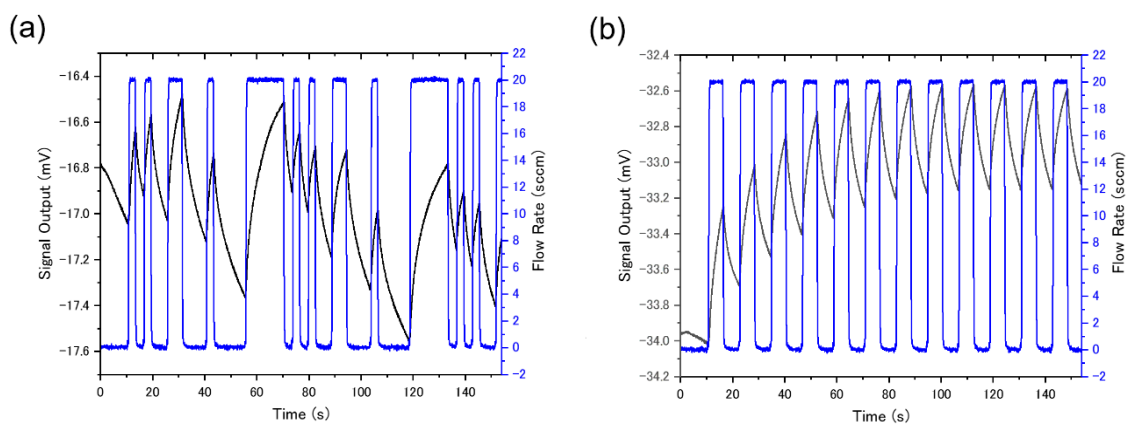

**Figure 7** Sensing signals from channel 1 of MSS Chip II and the monitored flow rates of a sample gas (ethanol) for (a) the m-sequence pseudorandom sequence and (b) the rectangular sequence.

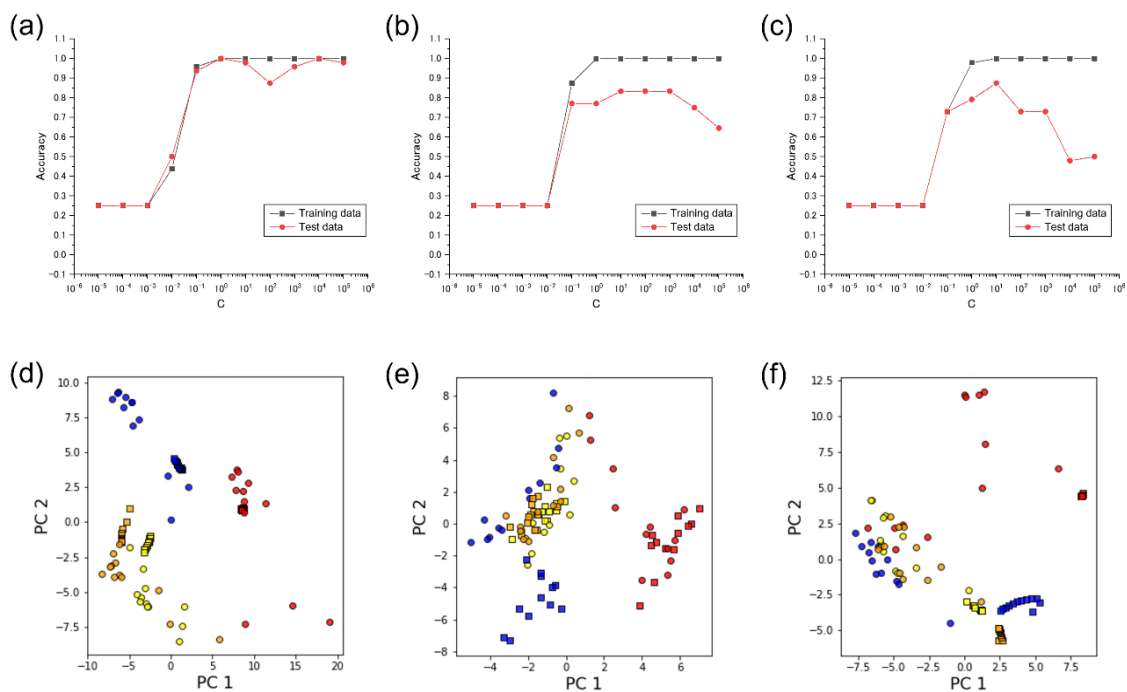

**Figure 8** Accuracies for training and test datasets plotted against regularization parameter $C$; the datasets were generated through the analysis methods based on (a) TFR, (b) AR model, and (c) FFT. Scatter plots of PCA from the datasets based on (d) TFR, (e) AR model, and (f) FFT. Yellow, red, blue, and orange correspond to ethanol, water, heptane, and ethyl acetate, respectively. Circle and square markers represent the m-sequence pseudorandom sequence and the rectangular sequence, respectively.

To compare the TFR-based analysis with other typical analytical methods, we also developed two models based on AR models and FFT. The results are also shown in Figure 8b-c. The suppressed accuracies in the test data for these analysis methods are attributed to the intrinsic difference from the TFR-based analysis, which is independent from the gas input pattern. Since the conventional analytical methods including AR models and FFT depend on gas input patterns, classification models which can be adapted to the different gas input patterns could not be developed. Accordingly, these approaches inevitably result in low accuracies for non-trained gas input patterns, which are common for practical measurements without gas flow control. The results of PCA on each dataset also indicate the robustness of TFR to gas input patterns. Figure 8d-f show the PCA scatter plots of the TFR-, AR-, and FFT-datasets, respectively. The color of the markers represents the solvents; yellow, red, blue, and orange correspond to ethanol, water, heptane, and ethyl acetate, respectively. The circle and square markers are the data obtained from the m-sequence pseudorandom sequence and the rectangular sequence, respectively. While there is a slight overlap between ethanol and ethyl acetate, TFR-dataset exhibits the highest separation even though the data obtained with the two different gas input patterns are mixed.

Conclusion

In this study, we have proposed a new gas identification protocol that does not require any gas flow controls. The key figure of this protocol is the data analysis method based on the TFR, which is independent of gas flow control and gas flow system. As TFR can be estimated only from the sensing signals of a gas sensor array (or a multichannel gas sensor chip), gas species can be identified without gas flow control units including pumps and MFCs. Combined with a miniaturized gas sensor chip, this gas identification protocol realizes the free-hand measurement, in which samples are measured simply by holding the sensor chip near the sample. To demonstrate the gas identification through the free-hand measurement, we developed a compact measurement system based on MSS and developed machine learning models for gas identification. By using the MSS chip coated with functionalized nano-/micro particles, we demonstrated the identification of not only solvent vapors but also odors of spices through the free-hand measurement with high accuracies. As MSS can utilize diverse materials for their receptor layers, the combination of the receptor materials in an MSS chip can be further optimized on the basis of the chemical composition of target odors. The robustness of the TFR-based analysis to gas input patterns was confirmed by building and validating classification models from measurement data obtained with explicitly different gas input patterns.

This gas identification protocol is not limited to MSS but applicable to any miniaturized gas sensor array or multichannel gas sensor chip. As gas flow control units such as pumps are not required in this protocol, a compact artificial olfactory system consisting of only a sensor chip and an electrical readout system can be realized, leading to the implementation of artificial olfaction in portable electronics and even in wearable devices. We believe this study will contribute to the realization of practical artificial olfaction.

Methods

Fabrication of MSS Chip I

The polymers were dissolved in solvents; poly(vinylidene fluoride), polysulfone and polycaprolactone were dissolved in N,N-dimethylformamide (DMF) while poly(4-methylstyrene) was dissolved in trichloroethylene. The concentration of the solutions was set at 1.0 g/L. The solutions are delivered onto the channels of MSS by an inkjet spotter (LaboJet-500SP, MICROJET Corporation).

Signal read-out system

The sensing signals of MSS are obtained as changes in the piezoresistors embedded in the four beams. The four piezoresistors form a Wheatstone bridge. To acquire the sensing signals from the MSS chip, a bridge voltage of -0.5 V was applied to the Wheatstone bridge circuits in the MSS chip with a digital-to-analog converter module (NI-9269, National Instruments). The sensing signals were collected with an analog-to-digital converters module (NI-9214, National Instruments). The sampling rate was set at

20 Hz.

Development of Machine learning models

From the obtained measurement data (i.e. 32 files of time-series data), we built machine learning models for gas identification based on TFR. The measurement data—the time-series data of 90 seconds—were divided by $t_m$; hence, the total number of the divided data becomes $90/t_m$. From each divided data, $K_{m,n}(f)$ was calculated according to Equation (5). Fast Fourier transform (FFT) was applied on each divided data, which was zero-meaned and multiplied by the Hann function in advance. Then, $K_{m,n}(f)$ was calculated for all the six combinations: $(m,n) = (1,2), (1,3), (1,4), (2,3), (2,4), (3,4)$. As the sampling rate and the time length of each data were 20 Hz and $t_m$ seconds, respectively, the components of $K_{m,n}(f)$ exist at 0, $1/t_m$, $2/t_m$, …, 10 Hz as complex numbers. Some frequency components were selected from each $K_{m,n}(f)$, and all the selected components were concatenated. In this study, $t_m$ was set at 3.0 seconds, and the frequency components ranging from 0.333 to 3.333 Hz were used. The complex number was divided into the absolute value and the argument. Therefore, the sample size and the dimension of the dataset are 960 and 120, respectively.

Based on the dataset of $K_{m,n}(f)$, we developed machine learning models for classifying the gas species. In this study, we employed six classifiers: support vector machines (SVMs) with a linear kernel, SVMs with a radial basis function (RBF) kernel, logistic regression (LR), decision trees (DTs), random forests (RFs), and multilayer perceptrons (MLPs). Machine learning models were optimized and validated through the grid search with the nested cross validation.

Acknowledgment

We thank Ms. Yuko Kameyama, Ms. Keiko Koda and Ms. Eri Sakon (CFSN, NIMS) for coating of the receptor layers. Keiko Koda also conducted the free-hand measurements. We also thank Ms. Takako Sugiyama (CFSN, NIMS) for synthesizing the nano-/micro-particles. This work was supported by the Leading Initiative for Excellent Young Researchers, Ministry of Education, Culture, Sports, Science and Technology (MEXT), Japan; a Grant-in-Aid for Young Scientists, 18K14133, MEXT, Japan; JST CREST (JPMJCR1665 and JPMJCR1666); a Grant-in-Aid for Scientific Research (A), 18H04168, MEXT, Japan; the Public/Private R&D Investment Strategic Expansion Program (PRISM), Cabinet Office, Japan; the Center for Functional Sensor & Actuator (CFSN), NIMS; the World Premier International Research Center Initiative (WPI) on Materials Nanoarchitectonics (MANA), NIMS; and the MSS alliance.